\documentclass[aps,pre,reprint,superscriptaddress,10pt]{revtex4-2}
\usepackage{amsmath,amssymb}
\usepackage{graphicx}% Include figure files
\usepackage{dcolumn}% Align table columns on decimal point
\usepackage{bm}% bold math
\usepackage{hyperref}
\usepackage{courier}
\usepackage{braket}
\usepackage{float}
\usepackage[usenames,dvipsnames]{xcolor}
\hypersetup{
  colorlinks   = true,	%Colours links instead of ugly boxes
  urlcolor     = blue, %Fuchsia,%Colour for external hyperlinks
  linkcolor    = blue, %Fuchsia, %Colour of internal links
  citecolor   = blue %DarkOrchid %Colour of citations#NavyBlue,Fuchsia
}
\newcommand{\cc}{\color{blue}}%use {\cc text text} command to change color: also \bf makes it bold
\begin{document}
\newcommand{\ve}{\varepsilon}
\title{Diverse coherence-resonance chimeras in coupled type-I excitable systems}
\author{Taniya Khatun}
\affiliation{Chaos and Complex Systems Research Laboratory, Department of Physics, University of Burdwan, Burdwan 713
  104, West Bengal, India}
\author{Biswabibek Bandyopadhyay}
\affiliation{Chaos and Complex Systems Research Laboratory, Department of Physics, University of Burdwan, Burdwan 713
  104, West Bengal, India}
\author{Tanmoy Banerjee}
\email[]{tbanerjee@phys.buruniv.ac.in}
%%%\homepage[]{Your web page}
\thanks{he/his/him}
\altaffiliation{}
\affiliation{Chaos and Complex Systems Research Laboratory, Department of Physics, University of Burdwan, Burdwan 713 104, West Bengal, India}
\date{\today}

\begin{abstract}
Coherence-resonance chimera was discovered in {\cc [Phys. Rev. Lett. 117, 014102 (2016)]}, which combines the effect of coherence resonance and classical chimeras in the presence of noise in a network of type-II excitable systems. However, the same in a network of type-I excitable units has not been observed yet. In this paper, for the first time, we report the occurrence of coherence-resonance chimera in coupled type-I excitable systems. We consider a paradigmatic model of type-I excitability, namely the saddle-node infinite period model and show that the coherence-resonance chimera appears over an optimum range of noise intensity. Moreover, we discover a unique chimera pattern that is a mixture of classical chimera and the coherence-resonance chimera. We support our results using quantitative measures and map them in parameter space. This study reveals that the coherence-resonance chimera is a general chimera pattern and  thus it deepens our understanding of role of noise in coupled excitable systems.
\end{abstract}

\maketitle

%##############################################################
%##################### (Introduction) #######################
%######################################
\section{Introduction}
Chimera patterns have been in the center of recent studies for the last two decades (see \cite{anna-book} and references therein). The inherent peculiarity of chimeras, such as coexistence of synchrony and asynchrony in networks of identical oscillators as the result of symmetry-breaking dynamics has intrigued the researchers in the field of natural and biological sciences \cite{chireview,schoell_rev}. Since its discovery in phase oscillators \cite{Y_Kuramoto_2002}, diverse chimera patterns are observed  \cite{G_C_Sethia_2008,amc_sethia,LaPeMa15,schoell-CD,Loos_2016,chaos_18,csod,poel-anna} and studied in several natural and man made systems \cite{lr16,schoell_qm,raj,keneth,dpll-chimera,lakh1,lakh2}. Apart from the academic interest, the relevance of chimera patterns in neuronal processes \cite{diba,diba2,hovel-moris,hra-hh} makes it a vibrant topic of research. Chimeras are found to be the underlying process that govern certain neuronal processes like unihemispheric sleep  \cite{Rattenborg_2000,Rattenborg_2006} and epileptic seizures  \cite{neuro1,neuro3}.  

In nature, noise and random fluctuations are inevitable \cite{vadim-book,anna-multiplex}. Noise manifests its most profound effects in excitable systems: noise-induced orders in the form of stochastic resonance \cite{sr} and coherence resonance \cite{fhn47,fhn48} have long been the subjects of intense research. As neurons are inherently excitable, these studies play crucial role in explaining several neuro-physiological processes such as noise-induced attractor switching that leads to seizure \cite{neuro2} (see \cite{sr-bio} and references therein). 
However, surprisingly, a little study has been done on the interplay of  noise and excitability in the context of chimera patterns. 
%The noise induced dynamical effects such as coherence resonance \cite{fhn48} have largely been untested in the context of chimeras. 
Only recently, Semenova et al. \cite{cr-chimera} reported a novel chimera pattern, called the coherence-resonance chimera (CR chimera) that combines temporal features of coherence resonance \cite{fhn48} and spatial properties of chimera states. In \cite{cr-chimera}, the authors considered a network of FitzHugh--Nagumo systems in their excitable state in the presence of noise and showed that, for an optimum range of noise strength, noise-induced spiking gives rise to spatial coherence--incoherence dynamics equivalent to chimera patterns. The FitzHugh--Nagumo model belongs to the type-II excitability, where the transition from excitable to oscillatory state occurs through Hopf bifurcation. In this context, there exist another broad class of excitable systems that exhibit type-I excitability, where the transition from excitable state to oscillation occurs through saddle-node bifurcation \cite{bard-book,iz-book}. Examples of type-I excitable systems include regular spiking neurons in rat somatosensory cortex \cite{example-type1-rat}, cerebellar stellate cells \cite{example-type1}, which are GABAergic interneurons found in the superficial molecular layer of the cerebellar cortex \cite{example-type1b}, and auditory nerve spike generators \cite{example-type1a} to name a few. However, surprisingly, the coherence-resonance chimera has not been observed yet in type-I excitable systems. In this paper we ask the long standing question: {\it do type-I excitable neurons exhibit coherence-resonance chimera?} If yes, what are the manifestations and origin of that chimera pattern?   

In this paper, for the first time, we show that coherence-resonance chimera is indeed exhibited by a network of type-I excitable systems. For our study we employ a paradigmatic model of type-I excitability, namely the SNIPER (saddle-node infinite period) model that gives limit cycle through saddle-node infinite period bifurcation, also known as saddle-node bifurcation on an invariant cycle \cite{bard-book}. We consider a network of nonlocally coupled identical SNIPER systems in their {\it excitable steady state} and show that the interplay of noise and coupling gives rise to coherence-resonance chimera pattern. 
% Unlike FHN systems, the width and height does not depend upon parameter value.
Unlike type-II excitable systems, we observe a unique chimera pattern in the moderate coupling strength which combines the features of classical chimera and coherence-resonance chimera. Using suitable measures we characterize the dynamical behaviors and delineate the dynamical zones in the parameter space. This study will show that the CR chimera is indeed a general chimera pattern and has other forms of manifestation.

%=============================================================
\section{Mathematical model}
\label{sec:snic}
%================single_sniper======================

We consider a network of $N$ identical type-I excitable systems obeying SNIPER model coupled through a nonlocal matrix coupling. The mathematical model of the network reads
\begin{eqnarray}
 \begin{split}\label{snic}
	\dot{x}_{i} &= x_{i}(1-x_{i}^2-y_{i}^2)+y_{i}(x_{i}-b) \\
	&\quad+\frac{\epsilon}{2P}\sum_{j=i-P}^{i+P}[b_{xx}(x_{j}-x_{i})+b_{xy}(y_{j}-y_{i})],\\
	\dot{y}_{i} &= y_{i}(1-x_{i}^2-y_{i}^2)-x_{i}(x_{i}-b)\\
	&\quad+\frac{\epsilon}{2P}\sum_{j=i-P}^{i+P}[b_{yx}(x_{j}-x_{i})+b_{yy}(y_{j}-y_{i})]+\sqrt{2D}\xi_{i}(t),            
\label{sniper}
 \end{split}
	\end{eqnarray}
where, $i=1,2,3...N$, $\epsilon>0$, is the coupling strength, and $P\in[1,N/2]$ is the number of nearest neighbors of each oscillator on either side. The limits $P=1$ and $P=N/2$ give the nearest neighbors and all to all coupling, respectively. $b$ is the bifurcation parameter ($b \in \mathbb{R}$). Here $\xi_{i}(t)\in \mathbb{R}$ is the normalized Gaussian white noise, i.e., $\langle\xi_{i}(t)\rangle=0$ and $\langle\xi_{i}(t)\xi_{j}(t')\rangle=\delta_{ij}\delta{(t-t')}$ $\forall i,j$ and  $D$ represents the noise intensity. 
The coefficients of $b_{lm}$, where $l,m \in [x,y]$, are the elements of the rotational matrix:
\begin{align}
\textbf{B}=
\quad
\begin{bmatrix}
b_{xx}& b_{xy} \\
b_{yx} & b_{yy}\\
\end{bmatrix}
=
\begin{bmatrix}
\cos{\phi}& \sin{\phi} \\
-\sin{\phi} & \cos{\phi}\\
\end{bmatrix},
\end{align}
where $\phi \in [-\pi,\pi]$. The matrix $\textbf{B} $ governs both direct coupling as well as cross coupling between $x$ and $y$ \cite{sniper25}. In this work we take $\phi=\pi/2-0.1$ as was prescribed in Refs.~\cite{sniper25,njp,cr-chimera}.

\begin{figure}
\centering
\includegraphics[width=0.48\textwidth]{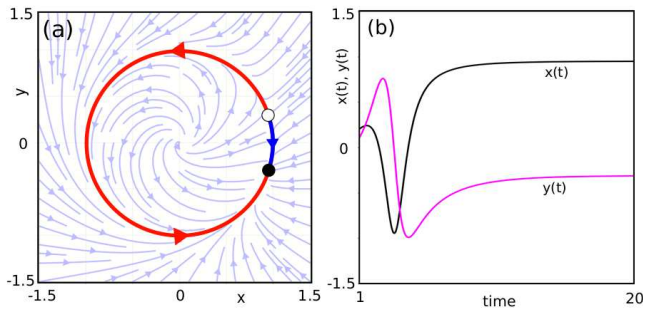}
\caption{A single SNIPER system in the excitable regime $b=0.95$. (a) Phase space diagram. Hollow circle (filled circle) denotes saddle point (stable node). The origin is an unstable focus. Two unstable manifolds approach the stable node along the periphery of a unit circle in two different directions. (b) Time series of $x(t)$ and $y(t)$.}
\label{f:snic}
\end{figure}

For $b<1$, the uncoupled system of Eq.~\ref{snic} has three fixed points: an unstable focus at the origin and a pair consisting of a saddle point and a stable node on the unit limit circle with coordinates $(b,+\sqrt{(1-b^2)})$, and $(b, -\sqrt{(1-b^2)})$, respectively. The dynamics is shown in Fig.~\ref{f:snic} for $b=0.95$ in phase space (a) and time series (b). At $b_{c}=1$ the saddle point and the stable node collide with each other through a saddle-node infinite period bifurcation (SNIPER) \cite{bard-book} and gives rise to a stable limit cycle. V\"{u}llings et al.  \cite{njp} studied this SNIPER model in the {\it oscillatory} zone (i.e. for $b>b_{c}$) without noise and observed clustered classical chimera patterns. 
However, in this work, our region of interest is $b<1$, i.e., all the SNIPERs are in the {\it non-oscillatory excitable steady state} and investigate the effect of noise on the collective dynamics of the coupled network. 

\section{Results}
We consider $N=1000$ SNIPER systems given by Eq.~(\ref{snic}) with $b<1$. In the absence of external noise, the individual nodes do not fire and the whole network stays in a stable homogeneous steady state. We fixed the bifurcation parameter at $b=0.995$, i.e., near the bifurcation point. We choose the phase-antiphase initial condition that has been widely used in the literature of chimera \cite{anna-book,schoell-CD,tan-CD}: $x_{(1-500)},y_{(1-500)}=1,-1$; $x_{(501-1000)},y_{(501-1000)}=-1,1$. However, we verify that for random initial conditions distributed on a unit circle ($x_{i}^2+y_{i}^2=1$) the network gives qualitatively the similar results.

%==========noise effects=================
\subsection{Coherence-resonance chimera}

%==========local order parameter==================
To understand the spatial coherence and incoherence dynamics of the chimera pattern, we use the local order parameter\cite{{fhn15},{fhn79}} that is defined as
\begin{align}
Z_{i}=\left|\frac{1}{2\delta_{m}}\sum_{|i-k|\leq \delta_{m}}e^{j\Theta_{k}}\right|,
\end{align}
where $j=\sqrt{-1}$, $i=1,2,...N$, and $\delta_{m}$ is the nearest neighbors of the $i$-th node on both sides. The geometric phase of the $i$-th element is defined by $\Theta_{i}=\arctan(y_{i}/x_{i})$\cite{sniper25}. The local order parameter $Z_{i}\approx{1}$ denotes the $i$-th oscillator belongs to the coherent group of the chimera pattern; $Z_{i}<1$, indicates the $i$-th oscillator belongs to the incoherent group \cite{cr-chimera,tbpre}. Here in our computation of local order parameter, we take the number of nearest neighbors $\delta_{m}=25$ \cite{cr-chimera}.

\begin{figure}[t!]
\centering
\includegraphics[width=0.5\textwidth]{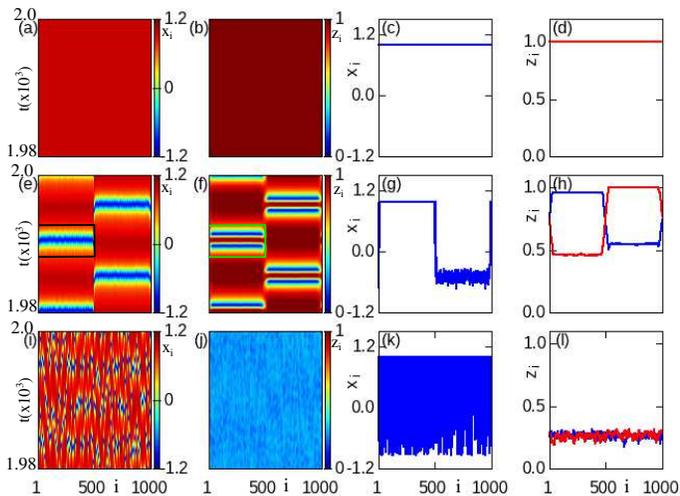}
\caption{Space-time plots of ($x_{i}$) (left column), local order parameter($Z_{i}$) (second column, $\delta_{m}=25$), snapshot of $x_{i}$ (third column) and snapshot of order parameter (right column) (at $t=1988$) for different noise intensities. (a,b,c,d) $D=0$: homogeneous steady state. (e,f,g,h) $D= 0.002$: coherence-resonance chimera. The rectangle containing the spatial incoherent nodes serves as a visual guidance. (h) shows snapshots of $Z_i$ at $t=1988$ (blue) and $t=1992$ (red) (i,j,k,l) $D=0.05$: incoherent in space and time. Parameter values: $\epsilon=0.33$, $b=0.995$,  $P=490$ and $N=1000$.}
\label{p490_multi}
\end{figure}
%=====================d=0.006======================
To demonstrate the effect of noise, we vary the noise intensity ($D$) for a fixed coupling strength ($\epsilon$) and coupling range ($P$): three distinct patterns are observed as shown in Fig.~\ref{p490_multi} using spatiotemporal plot of the variable $x_{i}$ and the corresponding local order parameter $Z_{i}$ (shown in the first two columns of Fig.~\ref{p490_multi}, respectively). For zero to a certain low noise intensity ($D\leq 0.00005$) the network stays in a homogeneous steady state. This is shown in Fig.~\ref{p490_multi} (a--d): the snap shots of $x_i$ and the corresponding local order parameter $Z_i$ in Fig.~\ref{p490_multi}(c) and Fig.~\ref{p490_multi}(d), respectively support this fact.
In the intermediate range of noise intensity ($0.00005\leq D \leq 0.008$) we observe coherence-resonance chimera (CR chimera), where certain oscillators form the spatial incoherent pattern and the remaining oscillators are in coherent motion with respect to each other. Figure~\ref{p490_multi} (e--h) depict the occurrence of CR chimera at an exemplary value of $D= 0.002$. The spatiotemporal plot of $x_i$ in Fig.~\ref{p490_multi}(e) shows that with time, the coherent and incoherent domain swap their position in space. However, in time domain they appear periodically. This is similar to the CR chimera defined for the type-II excitable system in Ref.~\cite{cr-chimera}. Interestingly, the interchange of coherent-inherent pattern has a resemblance with the noise-induced attractor switching in neuronal networks that leads to epileptic seizure \cite{neuro2}. Figure~\ref{p490_multi}(g) shows the snapshot of $x_i$ at $t=1988$ that clearly demonstrate that the oscillators of the right half are in incoherent motion while the oscillators in the left half are synchronized. Fig.~\ref{p490_multi}(h) (in blue) supports the observation through the corresponding snapshot of $Z_i$; it also shows the same (in red) in a later time ($t=1992$) where the coherent-incoherent zones are swapped.   
We find that, unlike \cite{cr-chimera}, the height of the incoherent domain does not depend upon the choice of the noise intensity or other coupling parameters.  
%This chimera effect can be understood with the help of a spatiotemporal plot, we see out of 1000 oscillators half of the oscillators are incoherent in space, and the left of the oscillators is coherent. 
Further increase in noise intensity leads to a complete incoherent dynamics in space and time. Unlike \cite{cr-chimera} the network does not exhibit a state of complete spatial incoherence and temporal coherence.  Fig.~\ref{p490_multi}(i-l) demonstrate the spatiotemporal incoherent dynamics for $D=0.05$. The snapshot of $x_i$ shows that all the oscillators are oscillating in an incoherent manner, which is also supported by the corresponding plot of $Z_i$ [Fig.~\ref{p490_multi}(l)].

\begin{figure}[t!]
\centering
\includegraphics[width=0.48\textwidth]{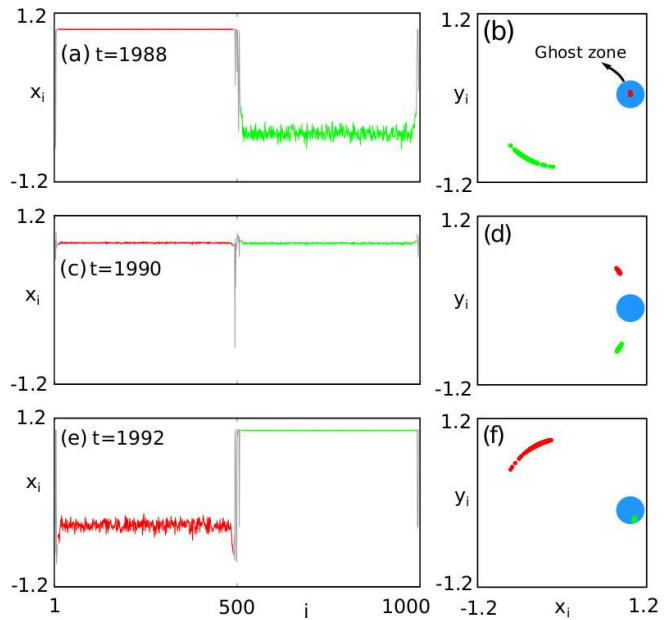}
\caption{Snapshots of $x_i$ and the corresponding phase space diagrams of coherence-resonance chimera of Fig.~\ref{p490_multi}(e) at (a,b) $t=1988$, (c,d) $t=1990$, (e,f) $t=1992$. The circular blob in the right-middle region of (b,d,f) sketch the ``ghost zone" where the dynamics become slower. Parameters are $D=0.002$, $\epsilon=0.33$, $b=0.995$, $P=490$, and $N=1000$.}
\label{snapshot}
\end{figure}
%===========d=0.006====================
To visualize the evolution of coherence-resonance chimera we inspect snapshots of the variable $x_{i}$ and the corresponding phase space in Fig.~\ref{snapshot}. Also, the genesis of the coherence-resonance chimera can be explained intuitatively from the phase space geometry of the SNIPER model. Near the bifurcation point, even when noise induced oscillations appear, a ``ghost zone" \cite{bard-book} [shown in circular shade in Fig.~\ref{snapshot}(b,d,f)] exists around $(x,y)=(b,0)$ that slows down the dynamics. Away from this ghost zone, the dynamics become faster. Fig.~\ref{snapshot}(a) shows that the oscillators of the left half (except a few in the boundaries) are spatially synchronized and the oscillators in the right half shows noise induced spatial incoherence at $t=1988$. The corresponding phase space diagram [Fig.~\ref{snapshot}(b)] ensures that the coherent oscillators are localized around a point in the phase space (in red), inside the ghost zone. Whereas the oscillators away from the ghost zone are spreading over a large region on the unit circle giving rise to the incoherent domain (in green). As time evolves, the incoherent oscillators are moving fast anti-clock wise and approach towards the ghost zone, whereas the coherent oscillators (in red) start moving away from the ghost zone. Figure~\ref{snapshot}(c,d) demonstrate the snapshot at  $t=1990$ where all the oscillators attain almost the same $x$ value. In a later time, the oscillators from the left half (in red) move faster and form the incoherent domain, whereas, the oscillators from the right hand side (in green) now move slowly and become coherent. This situation is shown in Fig.~\ref{snapshot}(e,f) at $t=1992$. The whole scenario repeats periodically in time and gives rise to alternatively switched coherence-resonance chimera. The genesis of the CR chimera is somewhat different from the CR chimera of FitzHugh--Nagumo model observed in \cite{cr-chimera} as there exist two slow zones separated by two fast zones in the phase space. Further, the CR chimera observed here slightly differs from that of \cite{cr-chimera}, as here we observed a few ``solitory" nodes that do not belong to spatial coherent-incoherent domain (see gray zone near $i=500$ in Figure~\ref{snapshot}(a,c,e)). In the next section we will show that, in a broad parameter zone these nodes give rise to a hybrid version of CR chimera.     

%====================snapshot_480====================
\subsection{Hybrid coherence-resonance chimera}
\begin{figure}
\centering
\includegraphics[width=0.48\textwidth]{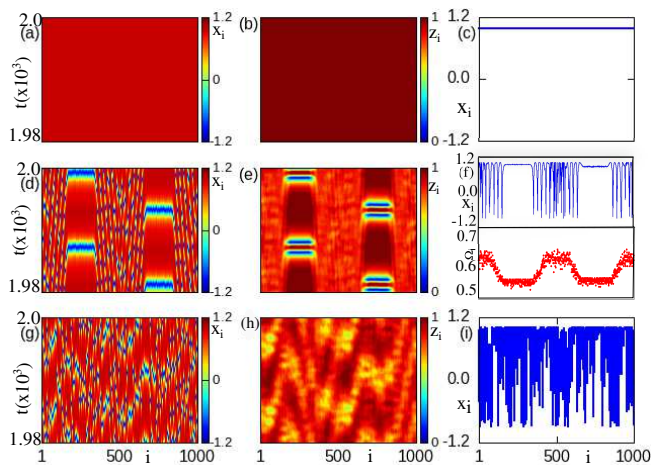}
\caption{Space-time plots of ($x_{i}$) (left column), local order parameter ($Z_{i}$) (second column, $\delta_{m}=25$) for different noise intensities. (a,b,c) $D=0$: steady state; (c) represents the snapshot of $x_i$. (d,e,f) $D= 0.002$, hybrid coherence-resonance chimera; upper panel of (f) gives the snapshot of $x_i$ at $t=1982$, the lower panel shows the corresponding mean phase velocity. (g,h,i) $D=0.05$ incoherent in space and time. Parameter values: $\epsilon=0.33$, $b=0.995$, $P=480$ and $N=1000$.}
\label{hcr}
\end{figure}

Apart from coherence-resonance chimera (CR chimera), we also observed a hybrid version of it, which manifests the signature of both the classical chimera and CR chimera. This chimera pattern appears in a broad parameter regime for comparatively a lower coupling range or stronger coupling strength. Figure.~\ref{hcr} demonstrate the scenario for $P=480$ and $\epsilon=0.33$ (the first and second columns show the spatiotemporal pattern of $x_i$ and the local order parameter $Z_i$, respectively). For a lower noise intensity, as before, the network stays in a state of homogeneous steady state [Fig.~\ref{hcr}(a,b,c)]. In an optimum range of noise we observe the {\it hybrid coherence-resonance chimera}, as shown in Fig.~\ref{hcr}(d): here the {\it spatial} coherent-incoherent domains of the CR chimera are separated by {\it spatiotemporal} incoherent oscillations. The corresponding local order parameter in Fig.~\ref{hcr}(e) supports this fact. Figure~\ref{hcr}(f) (upper panel) shows the snapshot of $x_i$ at $t=1982$, which clearly exhibits that the incoherent oscillations appear in between the coherent and incoherent domain of the original CR chimera. Therefore, in the hybrid CR chimera state the incoherent domain consist of two different kind of oscillatory dynamics: one corresponds to the {\it spatial incoherency} and the other to the {\it spatiotemporal incoherency}. We compute the mean phase velocity to characterize this chimera state. The mean-phase velocity \cite{anna-book} profile of each oscillator is given by
\begin{equation}
\Omega_{i}=\frac{2\pi M_{i}}{\Delta{T}},
\label{omega}
\end{equation}
where $M_{i}$ denotes the numbers of periods of the $i$-th oscillator in the time interval $\Delta{T}$. From the mean phase velocity profile of Fig.~\ref{hcr}(f) (lower panel) we can see that the spatiotemporal incoherent region shows an arc like shape (noisy due to the presence of noise) that is a signature of classical chimera. The hybrid CR chimera gets destroyed beyond a certain noise intensity: Fig.~\ref{hcr}(g,e,f) demonstrate the spatiotemporal incoherent dynamics for $D=0.05$. 

\begin{figure}
\centering
\includegraphics[width=0.48\textwidth]{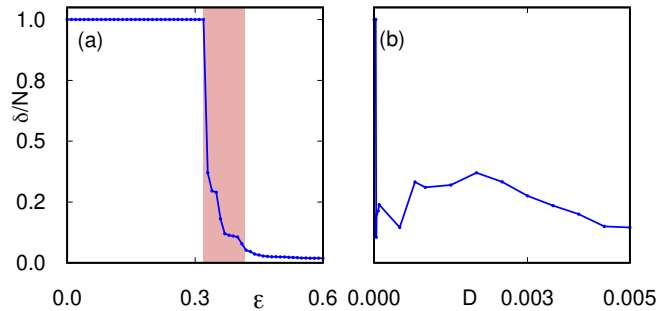}
\caption{ (a) Normalized length of coherent zone ($\delta/N$) vs. coupling strength ($\epsilon$) for $D=0.002$; The hybrid CR chimera appears in the shaded zone. (b) $\delta/N$ vs. noise intensity $D$ for $\epsilon=0.33$. Other parameters are $b=0.995$,  $P=480$ and $N=1000$.}
\label{delta}
\end{figure}

The spatial spreading of coherent and incoherent zone of hybrid CR chimera depends upon coupling strength ($\epsilon$) and noise intensity ($D$). We compute the variation of the normalized length of the coherent domain ($\delta/N$) with $\epsilon$ and $D$. Fig.~\ref{delta}(a) shows the plot of $\epsilon-\delta/N$ for $D=0.002$: the shaded zone represents the zone of occurrence of hybrid CR chimera. Left to this zone contains homogeneous steady state, and the right zone contains spatiotemporal incoherence. Fig.~\ref{delta}(b) gives the variation of $\delta/N$ with the noise intensity $D$ for a fixed coupling strength ($\epsilon=0.33$) and range ($P=480$). It shows that for $D=0$, all the nodes are in the steady state (i.e., $\delta/N=1$). As we increase the noise intensity, at first only a few nodes form the coherent domain, however, at an optimum range of noise intensity, larger number of nodes cooperate to form the coherent domain. Beyond that noise intensity, the number of coherent nodes gets reduced and ultimately all the nodes become incoherent giving spatiotemporal incoherent behavior. The presence of an optimum range of noise intensity is the signature of the coherence-resonance phenomenon.

\begin{figure}
\centering
\includegraphics[width=0.4\textwidth]{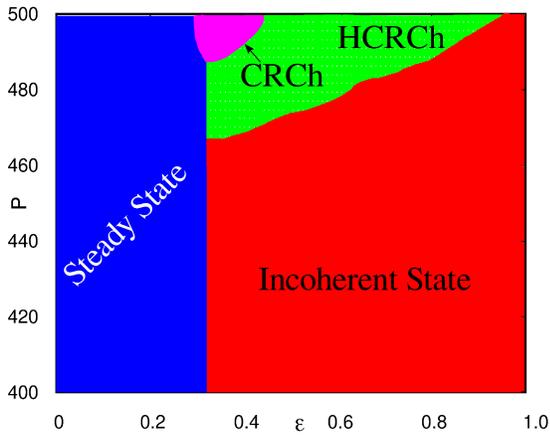}
\caption{Two parameter plot in $\epsilon-P$ space. CRCh: Coherence-resonance chimera, HCRCh: hybrid coherence-resonance chimera. Other parameters are $b=0.995$, $D=0.002$, and $N=1000$.}
\label{f:twop}
\end{figure}

Finally, we map all the dynamical behaviors  in the phase diagram of $\epsilon-P$ parameter space. Figure~\ref{f:twop} shows the two-parameter phase diagram for a constant noise level $D=0.002$. From the figure it can be noticed that up to $P\approx 470$ only two types of dynamics are possible: for the lower coupling strength all the nodes stay in the homogeneous steady state, however, for $\epsilon \ge 0.34$, incoherent oscillations set up in the network. Chimera patterns appear for $P\ge 470$ and $\epsilon \ge 0.32$. The CR chimera occurs near global coupling (i.e.,  $P\ge 485$) and the rest of the chimera patterns are hybrid CR chimeras. The boundaries of the incoherent state, the hybrid CR chimera, and CR chimera are multistable (not shown here). In fact, this hybrid CR chimera is found to be the most abundant chimera pattern in the network. For a random initial condition a pure CR chimera is not observed and the only chimera pattern is the hybrid CR chimera (see Appendix~\ref{app}).

%===============conclusion====================
\section{Conclusion}
In this paper, for the first time, we have shown that coupled type-I excitable systems exhibit coherence-resonance chimeras. Further, we have discovered a hybrid chimera pattern that exhibits the signature of both classical chimera and the coherence-resonance chimera, which occurs in a broad parameter region. We have considered excitable SNIPER system, which is a paradigmatic model of type-I excitability and showed that in a network an intermediate noise intensity breaks the homogeneity of the network and induces the coexisting spatial synchrony-asynchrony pattern, which is the combined manifestation of coherence resonance and chimera states. We have characterized all the chimera patterns with suitable measures and map them in the parameter space. Subsequently, we have explored the intuitive connection between the slow-fast dynamics and the observed chimera pattern.

The following are the results and observations that are new and unique in comparison to the type-II excitable system \cite{cr-chimera}:
(i) We have discovered a new chimera pattern, namely the hybrid coherence resonance chimera that is a mixture of classical chimera and the coherence-resonance chimera. This chimera pattern is unique in the sense that it consists of two ``dissipative structures" \cite{goldprigo}: namely the spatial coherence-incoherence pattern and the spatiotemporal incoherence pattern. (ii) With the random initial condition the only chimera pattern is the hybrid coherence resonance chimera. Therefore, in type-I excitable systems the hybrid coherence resonance chimera is the most abundant chimera pattern. (iii)~We have also explained the origin of the coherence  resonance chimeras; unlike the type-II excitable model \cite{cr-chimera} here the ghost region governed by the saddle-node bifurcation is responsible for the chimera pattern.

%This hybrid coherence-resonance chimera pattern was not observed in the network of type-II excitable system reported in \cite{cr-chimera}. 
%We have characterized all the chimera patterns with suitable measures and map them in the parameter space. Subsequently, we have explored the intuitive connection between the slow-fast dynamics and the observed chimera pattern. 
This study established that the notion of coherence-resonance chimera is indeed general and much broader: i.e., it is exhibited by both type-I and type-II excitable systems and its manifestation has diverse spatiotemporal dynamics. We believe that, the present study will improve our insight on the emergent dynamics of excitable systems in the presence of noise.

\begin{acknowledgments}
T.~K. and B.~B. acknowledge the financial assistance from the University Grants Commission, India in the form of Senior Research Fellowship. T.~B. acknowledges the financial support from the Science and Engineering Research Board (SERB), Government of India, in the form of a Core Research Grant [CRG/2019/002632].
\end{acknowledgments}

\appendix

\section{Random initial condition}
\label{app}
\begin{figure}[t!]
\centering
\includegraphics[width=0.49\textwidth]{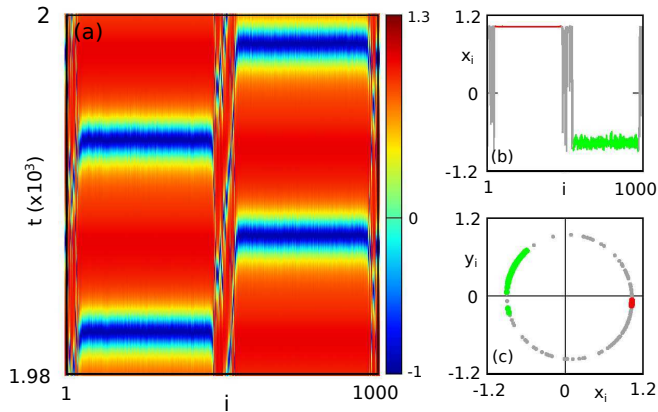}
\caption{Results for random initial condition on unit circle. (a) Spatiotemporal plot of $x_i$ showing hybrid CR chimera. (b) Snapshot of $x_i$ ($t=1990$). Gray nodes belong to the {\it spatiotemporal} incoherent domain. Nodes in green belong to the {\it spatial} incoherent zone of the hybrid CR chimera, whereas red nodes belong to the spatial coherent zone.  (c) The corresponding distribution of all the nodes in the phase space ($t=1990$). Other parameters are same as Fig.~\ref{p490_multi}(e), i.e., $\epsilon=0.33$, $b=0.995$, $D=0.002$, $P=490$ and $N=1000$.}
\label{rnd}
\end{figure}
We verify our results for a random initial condition distributed on a unit circle, $x_{i}^2+y_{i}^2=1$. We do not get any pure CR chimera for the random initial condition, instead hybrid CR chimeras are observed. Fig.~\ref{rnd}(a) demonstrates the spatiotemporal plot of the hybrid CR chimera with all the parameters same as Fig.~\ref{p490_multi}(e). It shows that the {\it spatial} coherent-incoherent domains are separated by {\it spatotemporal} incoherent domain. Fig.~\ref{rnd}(b) demonstrates the  snapshot of $x_i$ at $t=1990$. Here also one can see that the nodes from the spatial coherent domain (in red) are disconnected from spatial incoherent domain (in green) by a spatiotemporal dynamics which is essentially incoherent in nature (in gray). The corresponding phase space diagram of all the oscillators are plotted in Fig.~\ref{rnd}(c): the spatial incoherent nodes (in green) are spread but localized near the upper-left portion of the unit circle; spatial coherent nodes are localized near $(x_i,y_i)=(1,0)$, however, interestingly, the spatiotemporal incoherent nodes are not localized and spread all over the unit circle. We also compute the phase diagram which is similar to Fig.~\ref{f:twop} except here the CR chimera zone is replaced by the hybrid CR chimera pattern.

%\bibliography{cr}
\end{document}